\documentclass[reprint,amsmath,amssymb,pre,floatfix,nofootinbib]{revtex4-1} 
\usepackage{graphicx}
\usepackage{xcolor} 
\begin{document}

\title{Probing Lorentz Invariance Violation with Neutrino Factories}
\author{F.~Rossi-Torres}
\email{ftorres@ifi.unicamp.br}
\affiliation{Instituto de F\'isica Gleb Wataghin, Universidade Estadual de Campinas - UNICAMP, Rua S\'ergio Buarque de Holanda, 777, 13083-859, Campinas-SP, Brazil}

\begin{abstract}
In this article we show the modification of the number of neutrino events ($\nu_\mu+\bar\nu_\mu$) caused by Lorentz Invariant Violation (LIV), $\sigma=5\times 10^{-24}$ and $10^{-23}$, in neutrino oscillations for a neutrino factory at a distance of 7500 km. The momentum of the muons can vary from 10-50~GeV and we consider $2\times 10^{20}$ decays per year. The modifications in the number of events caused by this $\sigma$ LIV parameter could be a strong signal of new physics in a future neutrino factory.
\end{abstract}
\maketitle

\section{Introduction}

A certain type of neutrino beam is generated by a muon source of very high intensity.
These muons could be stored and allowed to decay in a ring containing a long straight
section that points in a desired direction. This kind of system was historically
called a {\it neutrino factory} and its main foundations were first developed by S. Geer~\cite{nufactseminal}.
The possible development of neutrino factories promises to increase the precision of neutrino oscillation parameters, assist in the determination of CP violation, and also determine the hierarchy of neutrinos. 

In Geer's seminal article it was pointed out that there are several improvements to neutrino beam 
experiments that could be achieved using neutrino factories. One improvement could be in the 
determination of the $\nu_e$ and $\nu_\mu$ fluxes since these are important sources 
of systematic errors. Also a more ``pure'' beam could be obtained, since the $\nu_\mu$ 
is contaminated by $\nu_e$ from $K^+$ three body decay for neutrino beam experiments. All of this ``contamination''
makes it very difficult for the precision experimentation in neutrino oscillation physics. 

Several proposals of neutrino factories have been made~\cite{nufact1,nufact2,nufact3,nufact4,nufact5,nufact6}, but its realization is still far from happening despite great effort and advances in research and development. Recent reviews on neutrino factories and their possible capabilities and physical potentials can be found in~\cite{nufactinterest1,nufactinterest2}. 

The principle of Lorentz invariance tells us that equations that describe some natural physical phenomena have the same structure in all reference frames. So, if we find some system with a violation of this principle, it would be evidence of new physics and the possibility for theoretical development in the understanding of the most basic physical laws. There are several proposals and tests involving Lorentz invariance violation (LIV)~\cite{perspective_lorentz1,perspective_lorentz2}. For a very recent review on tests of LIV in effective quantum field theories and the gravity sector, see~\cite{perspective_lorentz3}. Despite this fact, we do not have compelling evidence of LIV and its effects are supposed to be very small and suppressed by the Planck scale ($M_p \sim 10^{19}$~GeV). Currently, we have limits on LIV based on experiments with electrons, photons and neutrons, for example. Existing limits can be found in~\cite{table_liv}. Generally, these limits are put in a framework called the Standard-Model extension (SME)~\cite{sme1,sme2} which is an extension of the Standard Model with Lorentz violating terms in the Lagrangian. The recent global model for neutrino oscillations based on the SME has shown consistency with all compelling data from accelerator, atmospheric, reactor, and solar neutrino experiments and it even reproduces the anomalous low-energy excess observed in MiniBooNE \cite{sme3,sme4}.  

Neutrinos can play an important role in this LIV picture~\cite{neutrinos_lorentz1,neutrinos_lorentz2,neutrinos_lorentz3,neutrinos_lorentz4}. One of the ways of determining the LIV is by measuring the neutrino ($\nu$) time of flight~\cite{neutrinos_lorentz3}. For example, this measurement was done by the MINOS experiment~\cite{minos}, where they compared the detection times in the near and far detectors of 3~GeV neutrino beams. The result was $(v-c)/c=5.1\pm 2.9\times 10^{-5}$ at 68\% C.L. Also, in a recent analysis, the OPERA experiment determined $-1.8\times 10^{-6}<(v-c)/c<2.3\times 10^{-6}$ at 90\% C.L~\cite{opera}. Two other experiments recently determined the neutrino velocity: ICARUS with $\delta t = \delta t_c-\delta t_\nu = 0.10\pm 0.67\pm 2.39$~ns \cite{icarus} and LVD with $-3.8\times 10^{-6}<(v-c)/c<3.1\times 10^{-6}$ at 99\% C.L~\cite{lvd}.
 
Another way to investigate LIV is by making use of neutrino oscillations since there will be modifications in the energy-momentum relation of the particle and therefore its Hamiltonian will be modified~\cite{coleman}. Models of string theory and extra dimensions have also explored the LIV possibility by the breaking of CPT symmetry. If CPT is violated, then there will be differences in the oscillation probabilities of neutrinos compared with the ones of antineutrinos. This possibility was investigated in the context of neutrino factories in~\cite{cpt1,cpt2}. 

In this work we investigate the modification in the total detected number of events ($\nu_\mu+\bar\nu_\mu$) by the introduction of the LIV parameter in neutrino oscillations for a neutrino factory with $2\times 10^{20}$ muon decays per year localized at a distance of 7500~km from the detector. The muons are accelerated until they reach a momentum which varies from 10~GeV/c to 50~GeV/c. According to our calculations, modifications in the number of events can, in principle, happen when we consider the LIV parameter not equal zero since there is a modification in the oscillation pattern. For the oscillation to be modified by the Lorentz violating parameter, which we call $\sigma_i$, we must have $E\delta \sigma_{ij} L\approx \pi$. So taking a distance, $L$, of 7500~km, for example, and an energy, $E$, of about 10~GeV, $\delta \sigma_{ij}$ must be about $10^{-23}$. In this article we will choose $\delta \sigma_{ij}=1\times 10^{-23}$ and $\delta \sigma_{ij}=5\times 10^{-24}$ to compare with the standard situation: $\delta \sigma_{ij}=0$. The reason we do that choice is that they maximize the LIV effect for the baseline chosen and the neutrino factory energy range.

This article is organized as follows. In Sec.~\ref{production} we present the main principles of the production of neutrinos in a neutrino factory. In Sec.~\ref{propagation} and sec.~\ref{detection}, we describe the neutrino propagation and its interaction with matter on the Earth including the LIV parameter, culminating with its detection. In Sec.~\ref{results}, we present our results by showing the number of events for scenarios including the LIV parameter, $\sigma$, and also the standard situation with no Lorenz violation ($\sigma=0$). Finally, we conclude our work in Sec.~\ref{conclusions}.

\section{Neutrino production}\label{production}

As pointed out in~\cite{geer2}, neutrino factories have basic concepts enumerated as followed: (i) a pion source which is produced by a multi-GeV proton beam (produced by a multi-MW proton source) focused at a particular target; (ii) secondary charged pions radially confined by a high-field target solenoid; (iii) the production of positive and negative muons by the decay of pions in a long solenoidal channel; (iv) the capture of muons by rf cavities and reductions in the spreading of energy; and (v) a reduction of the transverse momentum of the muons in an ionization cooling channel. We are not going to discuss in detail the neutrino factory R\&D. For more information about this, see~\cite{nufactinterest1,nufactinterest2}. 

In a general neutrino factory, neutrinos are produced by $\mu^{\pm}$ decay. All flavors are produced in a neutrino factory, except $\nu_\tau$ and $\bar\nu_\tau$. We have the following decays: $\mu^-\to e^-+\bar\nu_e+\nu_\mu$ and $\mu^+\to e^++\nu_e+\bar\nu_\mu$. In the muon rest frame, we can write the distribution of $\bar\nu_\mu$ ($\nu_\mu$) in the following expression~\cite{gaisser,barger}:
\begin{equation}
\frac{d^2N_{\nu_\mu}}{dy d\Omega}=\frac{2y^2}{4\pi}[(3-2y)\mp (1-2y)\cos\theta],
\label{dist_nu_muon}
\end{equation} 
where $y\equiv 2E/m_\mu$, $E$ represents the neutrino energy, $m_\mu$ is the muon rest mass, and $\theta$ is the angle between the neutrino momentum vector and the muon spin direction. In Eq.~(\ref{dist_nu_e}), we show the expression for the distribution of $\nu_e$ ($\bar\nu_e$):
\begin{equation}
\frac{d^2N_{\nu_e}}{dy d\Omega}=\frac{12y^2}{4\pi}[(1-y)\mp (1-2y)\cos\theta].
\label{dist_nu_e}
\end{equation} 

For simplicity we are going to consider unpolarized muon beams, so Eqs.~(\ref{dist_nu_muon}) and (\ref{dist_nu_e}) become $\frac{d^2N_{\nu_\mu}}{dy d\Omega}=\frac{2y^2}{4\pi}(3-2y)$ and $\frac{d^2N_{\nu_e}}{dy d\Omega}=\frac{12y^2}{4\pi}(1-y)$. We will consider $2\times 10^{20}$ $\mu^\pm$ decays per year. The $\mu^\pm$ momentum in our neutrino factory will vary from 10~GeV/c to 50~GeV/c. In Ref.~\cite{tunnell} there was a proposition of a very low energy neutrino factory, where the muon energy is about 2-4~GeV and in~\cite{winter}, Sec.~III-IV, there is a discussion about the setup/optimization and the systematics of this low energy neutrino factory. A recent proposition of very low energy neutrino factory, with a initial beam energy of about 300~MeV, called MOMENT, was made in~\cite{china_proposition}.

\section{Neutrino Propagation}\label{propagation}

After the production, neutrinos will propagate through the Earth and will travel a distance ($L$) of about 7500~km, approximately the distance from Fermilab to Gran Sasso. We know that those neutrinos will oscillate during their propagation and will suffer MSW effects~\cite{msw}. This baseline $L$ is usually known as the ``magic'' baseline~\cite{magic} because it is the distance that, in principle, can help to solve the unknown neutrino mass ordering and improve, for example, the determination of $\theta_{13}$\footnote{In \cite{magic}, if there is a combination of this 7500~km ``magic'' baseline with another baseline of 3000~km, this set of baselines could be used also to determine a CP violation phase.}. For previous research on the interactions with the matter from Earth and its effects in very long baselines neutrino oscillations, such as neutrino factories, see~\cite{freund1,freund2}. The equation of evolution, in flavor basis and using natural units ($c=1$ and $h=1$), is described by
\begin{equation}
i\frac{d\Psi_\alpha}{dx}=H_f \Psi_\alpha,
\label{scho}
\end{equation}
where $\Psi_\alpha=(\Psi_{\alpha e}~\Psi_{\alpha \mu}~\Psi_{\alpha \tau})$, for $\alpha=e,\mu,\tau$, and $x$ is the neutrino position. The effective Hamiltonian ($H_{eff}\equiv H_f$) in this case includes the effects of matter from Earth, represented by the function $V(x)$, which contains the charged current interactions of neutrinos with the electrons in Earth:
\begin{equation}
V(x)=\sqrt{2} G_F n_e(x),
\label{pot}
\end{equation} 
where $G_F$ is the Fermi coupling constant ($G_F=1.166\times 10^{-5}$~GeV$^{-2}$) and $n_e(x)$ represents the electronic density, which depends on the neutrino position along its path. Considering $n_e(x)$ a spherically symmetric distribution, the core and the mantle are the two main parts, as opposed to the other parts such as the shells and other layers. The Earth's radius is about 6371~km and the core has a radius of 3486~km. The mantle has 2885~km of depth. The neutrino path through Earth is determined by the nadir angle ($\theta_n$). For $\theta_n\le 33.17^\circ$, or $L\ge 10660$~km, neutrinos will cross the core. Since we are considering distances less than this particular value, our neutrino path crosses only the mantle region, which has an average density of $\bar n_e^{mantle}\approx 2.2 N_A$~cm$^{-3}$~\cite{prem}, where $N_A$ is Avogadro's number. As it was developed in~\cite{con1,con2,con3}, because there is no significant change in the electronic density, we consider $n_e(x)$ constant along the neutrino path and equal to the average: $n_e(x)=\bar n_e^{mantle}$. From now on we are going to represent $V(x)\equiv \bar V$. For modifications in the oscillations using a non-constant profile see \cite{earth_profile}, where there are some changes in the oscillation probabilities, however they are only a few percent.

We can write $H_{eff}\equiv H_f=UHU^\dagger+\bar V$, where $U$ is the PMNS mixing matrix, characterized by four parameters: $\theta_{12}$, $\theta_{23}$, $\theta_{13}$,and $\delta$, the phase related to possible CP-violation. $H$ is the Hamiltonian in the mass eigenstates basis, $(\nu_1~\nu_2~\nu_3)$, and it is related to the LIV modification, because for a general mass eigenstate, $i$, we can write: 
\begin{equation}
E_i\approx (1+\sigma_i)E+\frac{m_i^2}{2E},
\label{livdisp}
\end{equation}
where we considered $m_i\ll p_i$, $m_i$ is the mass of each mass eigenstate, $p_i$ is their respective momentum, for $i=1,2,3$, and $E$ represents the neutrino energy. The LIV parameter is represented by $\sigma_i$ in the first order approximation. For $\sigma_i=0$, we recover the usual and standard dispersion relation for a neutrino in the ultra-relativistic approximation. This $\sigma$ will represent a small perturbation in the dispersion relation when we break the Lorentz invariance of some Lagrangian. In other words, LIV is produced during the propagation of the mass eigenstates, so each mass eigenstate will propagate with a different group velocity. Mixing angles remain equal to the standard neutrino oscillation scenario ($\sigma=0$) and no drastic deviation happens from the current status of neutrino oscillation, since current data does not point to any clear evidence of LIV. We also consider that, for simplicity, the LIV parameter acts in the same way for neutrinos and antineutrinos. For a different theoretical scenario and a more advanced discussion, see~\cite{greenberg}. In this article we do not show any Lagrangian with a Lorentz breaking term, but we invite the reader to look for a more theoretical motivation of this kind of phenomena in~\cite{coleman}, for example.

In order to give more meaning to our LIV parameter $\sigma$, we will compare it with a particular case of the SME discussed in section 4.2.1 of~\cite{neutrinos_lorentz2} and detailed in~\cite{newkost}. In the isotropic approximation (no Lorentz violation from rotations) and neglecting flavor information of the SME, the neutrino group velocity ($v_g^{sme}$) can be written as:
\begin{equation}
v_g^{sme}=1-\frac{m^2}{2E^2}+\mathaccent'27 c+\sum_{n=1}^\infty k^{(n)} E^n,
\label{jsdiaz}
\end{equation} 
for a generic mass $m$ and energy $E$. The quantities $k^{(n)}$ are proportional to the isotropic LIV coefficients of the SME and $\mathaccent'27 c=-4 c^{TT}/3$, where $c^{TT}$ is the coefficient for Lorentz violation related to the SME Lagrangian\footnote{They are tensor vaccum expectation values which are background fields that break the Lorentz symmetry.}. For simplicity, the following comparison will be done with $\mathaccent'27 c$.
 
If we consider our dispertion relation in Eq.~(\ref{livdisp}), the group velocity ($v_g^i=\left(\frac{dp}{dE_i}\right)^{-1}$), for each mass eigenstate, can be written for our LIV case as:
\begin{equation}
v_g^i\approx 1-\frac{m_i^2}{2E_i^2}+\frac{\sigma}{2}.
\label{mygroup}
\end{equation}
So, if we compare Eq.~(\ref{mygroup}) with Eq.~(\ref{jsdiaz}), $\sigma/2=\mathaccent'27 c$. Note that, by construction and assumptions, our $\sigma$ is not explicitly dependent on energy and, for this reason, we neglected the terms $k^{(n)}$ in $v_g^{sme}$ to do a fair comparison.

The Hamiltonian in the mass basis can be explicitly written as:
\begin{equation}
H=E+\left(
\begin{array}{ccc}
0 & 0 & 0\\
0 & E(\sigma_2-\sigma_1)+\frac{\Delta m^2_{21}}{2E} & 0\\
0 & 0 & E(\sigma_3-\sigma_1)+\frac{\Delta m^2_{31}}{2E}
\end{array}
\right).
\label{liv_matrix}
\end{equation}
For the distance considered and the energy range of the order of tens of GeV, we can ignore the CP violation effects~\cite{cp_violation_ignore,winter}. In Eq.~(\ref{liv_matrix}), we consider, for simplicity, that the LIV modifications are $\sigma_i-\sigma_j\equiv \sigma$. Also, $\Delta m^2_{ij}\equiv m_j^2-m_i^2$, for $j=1,2,3$. We can then consider the best-fit parameters of the oscillation parameters determined by~\cite{bari}. Based on the solar experiments and the KamLand reactor
$\bar\nu_e$, $\Delta m^2_{21}=7.54\times 10^{-5}$~eV$^2$ and $\sin^2\theta_{12}=0.307$ are 
the best-fit values. For normal hierarchy (NH), $\Delta m^2_{32}=2.43\times 10^{-3}$~eV$^2$, $\sin^2\theta_{23}=0.386$, and $\sin^2\theta_{13}=0.0241$. For inverted hierarchy (IH), we use the following oscillation parameters: $\Delta m^2_{32}=2.42\times 10^{-3}$~eV$^2$, $\sin^2\theta_{23}=0.392$, and $\sin^2\theta_{13}=0.0244$. These parameters are fully summarized in Table~\ref{tab_osc_par}.
\begin{table}[ht]
\centering
\begin{tabular}{cccccc}
\hline
&$\Delta m^2_{21}$~(eV$^2$) & $\Delta m^2_{31}$~(eV$^2$) & $\sin^2\theta_{21}$ & $\sin^2\theta_{23}$ & $\sin^2\theta_{13}$\\
\hline
NH & $7.54\times 10^{-5}$ & $2.43\times 10^{-3}$ & 0.307 & 0.386 & 0.0241 \\
\hline
IH & $7.54\times 10^{-5}$ & $2.42\times 10^{-3}$ & 0.307 & 0.392 & 0.0244\\
\hline
\end{tabular}
\caption{Values of the best-fit oscillation parameters extracted from the Bari group~\cite{bari}.}
\label{tab_osc_par}
\end{table}

We calculated the conversion probabilities for the so-called {\it golden channel}, $\nu_e\to\nu_\mu$ and $\bar\nu_e \to \bar\nu_\mu$. Also, we calculated the survival probabilities for the $\nu_\mu\to\nu_\mu$ and $\bar\nu_\mu\to\bar\nu_\mu$ channels. These calculations are justified because we are interested in events related to the productions of $\mu^\pm$ in the detector and only $\nu_\mu$ and $\bar\nu_\mu$ are going to generate these muons, as we explain in next section. The extraction of the probabilities for each neutrino energy, which varies from 1~GeV to 50~GeV, and for the fixed baseline of $L=7500$~km, is done by numerically solving the differential equation shown in Eq.~(\ref{scho}). The initial conditions for $\nu_e\to \nu_\mu$ (or $\bar\nu_e\to \bar\nu_\mu$) are $\Psi_e(0)=(\Psi_{ee}~\Psi_{e\mu}~\Psi_{e\tau})=(1~0~0)$. Conversely, for $\nu_\mu\to \nu_\mu$ (or $\bar\nu_\mu\to \bar\nu_\mu$) they are $\Psi_\mu(0)=(\Psi_{\mu e}~\Psi_{\mu\mu}~\Psi_{\mu\tau})=(0~1~0)$. 

We also consider the different effects of the ordering of the neutrino masses. We must remember that this question is still an open one in neutrino physics. NH is considered when we have the lightest mass eigenstates being $m_{\nu_1}$ and the heaviest, $m_{\nu_3}$: $m_{\nu_1}<m_{\nu_2}\ll m_{\nu_3}$. In the case of IH we have: $m_{\nu_3}\ll m_{\nu_2}< m_{\nu_1}$.   

\begin{figure}[h!tb]
\centering
\includegraphics[scale=0.3]{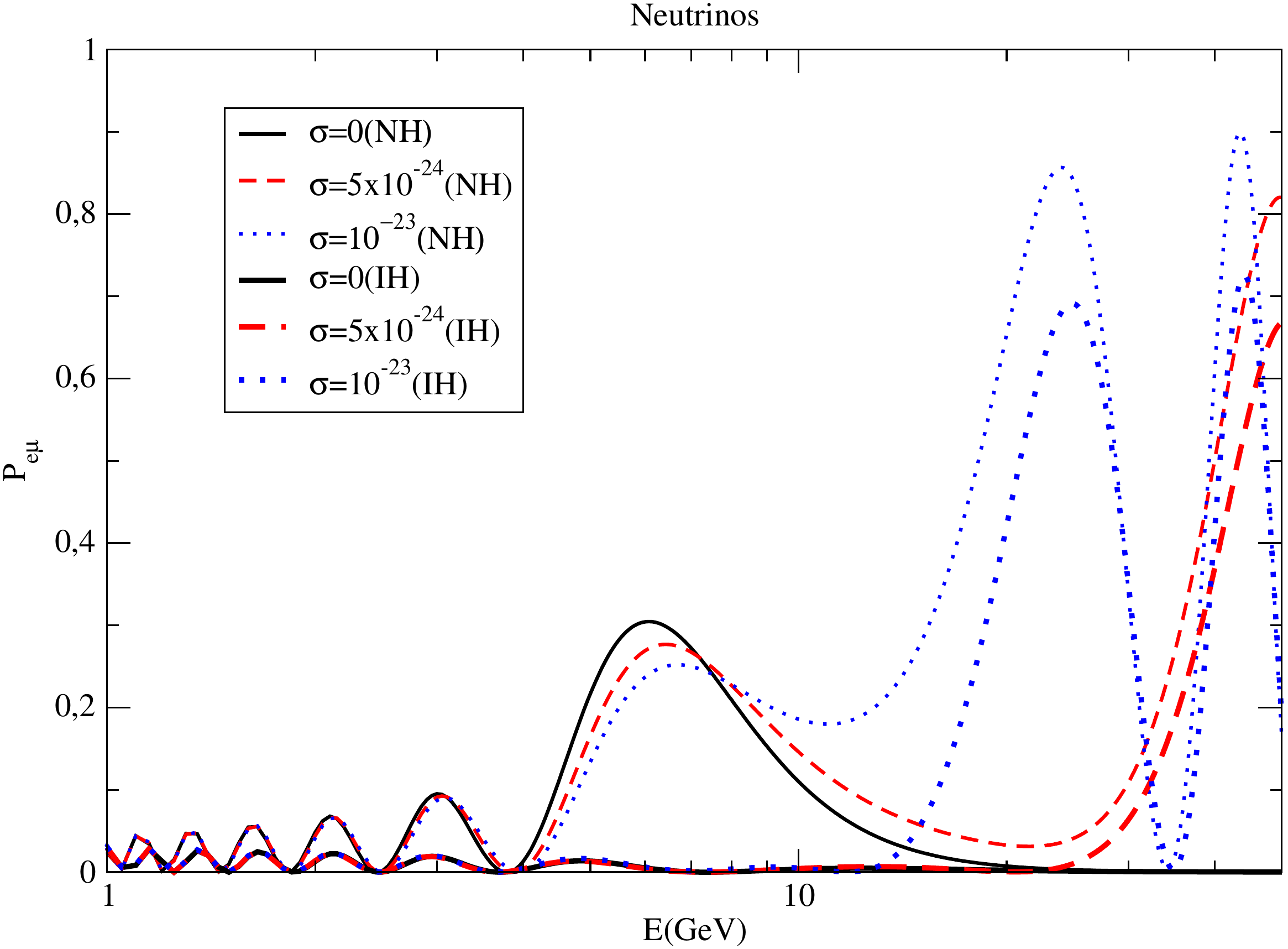}
\caption{In this figure we show the conversion probability for the {\it golden channel}, $\nu_e\to\nu_\mu$. We consider a baseline of $L=7500$~km. The solid curve is for the standard picture ($\sigma=0$) and the dashed and dotted curves are the LIV cases with $\sigma=5\times10^{-24}$ and $\sigma=10^{-23}$, respectively. The thinner curves are for NH and the thicker curves are for IH.}\label{fig1}
\end{figure}

\begin{figure}[h!tb]
\centering
\includegraphics[scale=0.3]{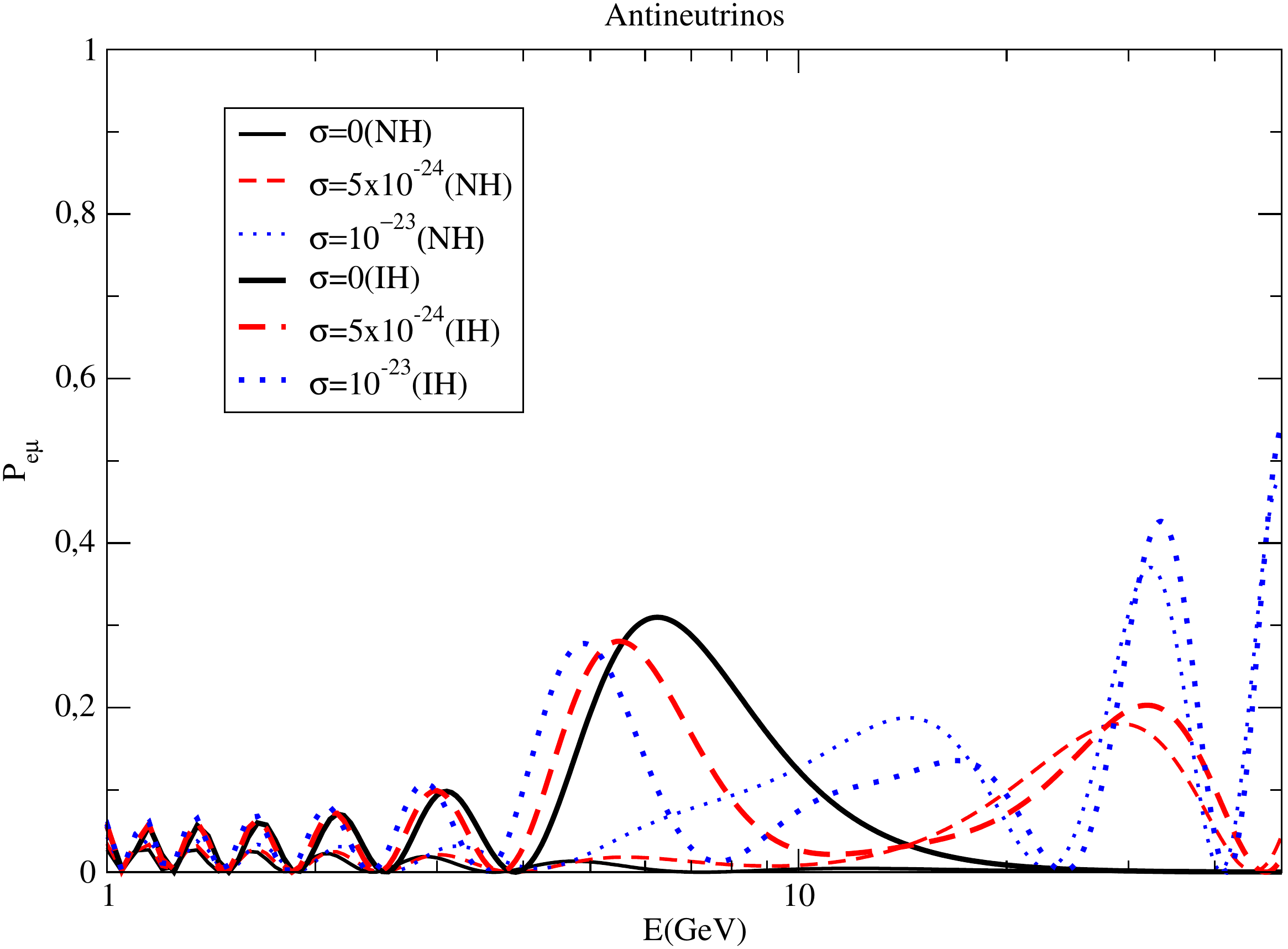}
\caption{In this figure we show the conversion probability for the {\it golden channel}, $\bar\nu_e\to\bar\nu_\mu$. We consider a baseline of $L=7500$~km. The solid curve is for the standard picture ($\sigma=0$) and the dashed and dotted curves are the LIV cases with $\sigma=5\times10^{-24}$ and $\sigma=10^{-23}$, respectively. The thinner curves are for NH and the thicker curves are for IH.}\label{fig2}
\end{figure}

In Fig.~\ref{fig1}, we plot the conversion probabilities for the channel $\nu_e\to\nu_\mu$. Solid curves represent the standard picture ($\sigma=0$) without LIV and dashed and dotted curves are the LIV cases for $\sigma=5\times10^{-24}$ and $\sigma=10^{-23}$, respectively. Thinner curves represent NH and thicker curves represent IH. First, for the standard case, we notice that the peak of the conversion probability for $\nu_e\to\nu_\mu$, in the NH case, occurs around $\sim 6.5$~GeV, which is the resonance energy ($E_{res}$) for the Earth matter potential ($V$), the $\theta_{13}$ angle, and the $\Delta m^2_{31}$ considered here: $E_{res}=\frac{\Delta m^2_{31} \cos 2\theta_{13}}{2 V}$. Notice that for the IH case, there is a suppression in the oscillation compared to the NH case, since interactions with matter are practically suppressed. In Fig.~\ref{fig2}, we have the same pattern, but now for the {\it golden channel}, $\bar\nu_e\to\bar\nu_\mu$. For approximately $E\ge 4$~GeV the presence of the LIV $\sigma$ factor significantly changes the values of the probabilities, as we can see in Fig.~\ref{fig1} and \ref{fig2}.  

When we take into account $\sigma$, the presence of a new oscillation term explains why these dotted and dashed curves have new oscillation peaks. There are modifications in the oscillation length and new resonances appear in the propagation. We stress the fact that for $\sigma \lesssim 1\times 10^{-24}$,
there is a coincidence between the LIV conversion probabilities with the standard situation ($\sigma=0$). Of course, in the present situation and analysis, this is a very weak limit since there was no simulation of experimental data done for the neutrino factory we are taking into account. 

\begin{figure}[h!tb]
\centering
\includegraphics[scale=0.3]{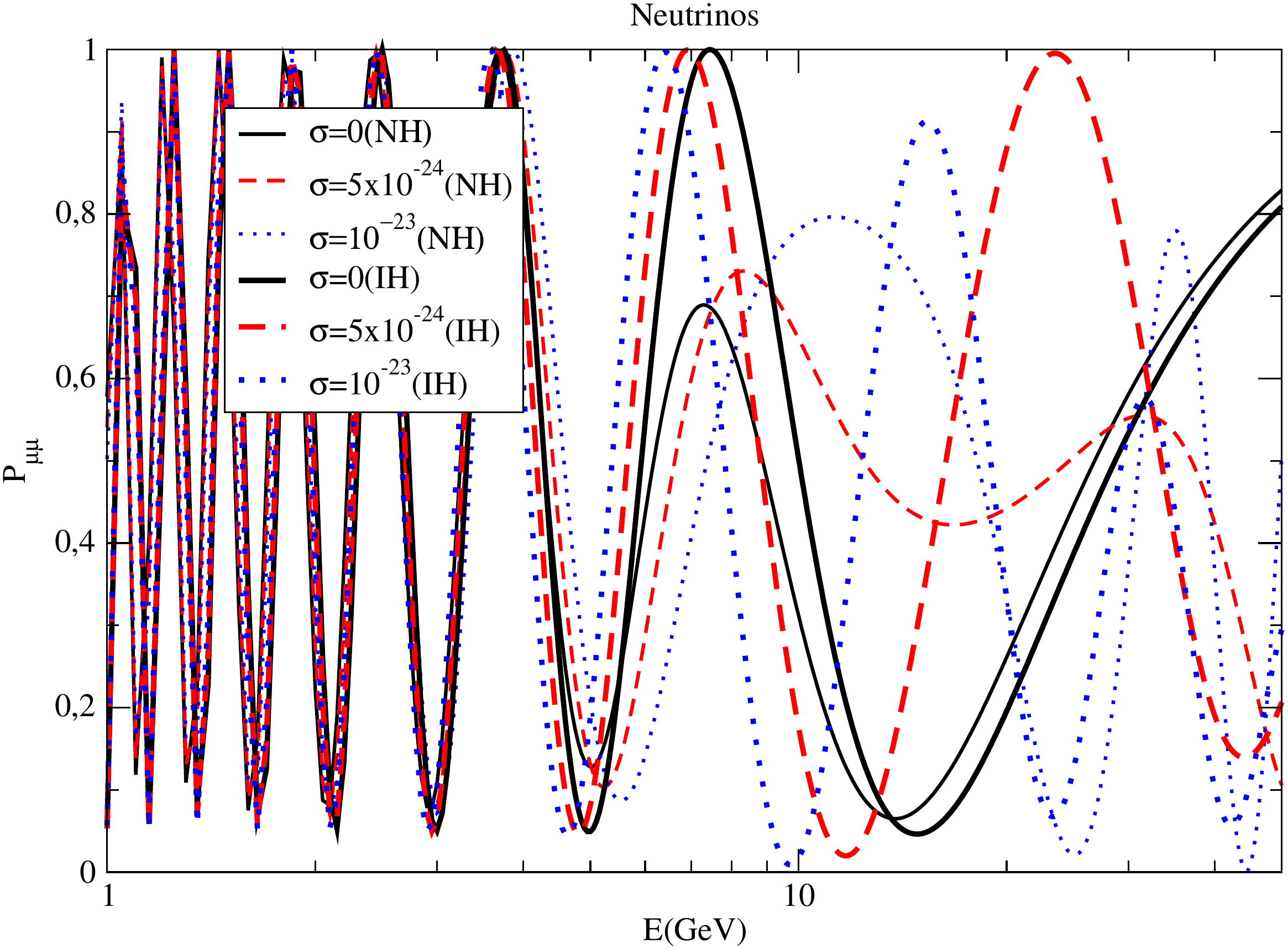}
\caption{In this figure we show the conversion probability for the process $\nu_\mu\to\nu_\mu$. We consider a baseline of $L=7500$~km. The solid curve is for the standard picture ($\sigma=0$) and the dashed and dotted curves are the LIV cases with $\sigma=5\times10^{-24}$ and $\sigma=10^{-23}$, respectively. The thinner curves are for NH and the thicker curves are for IH.}\label{fig3}
\end{figure}

\begin{figure}[h!tb]
\centering
\includegraphics[scale=0.3]{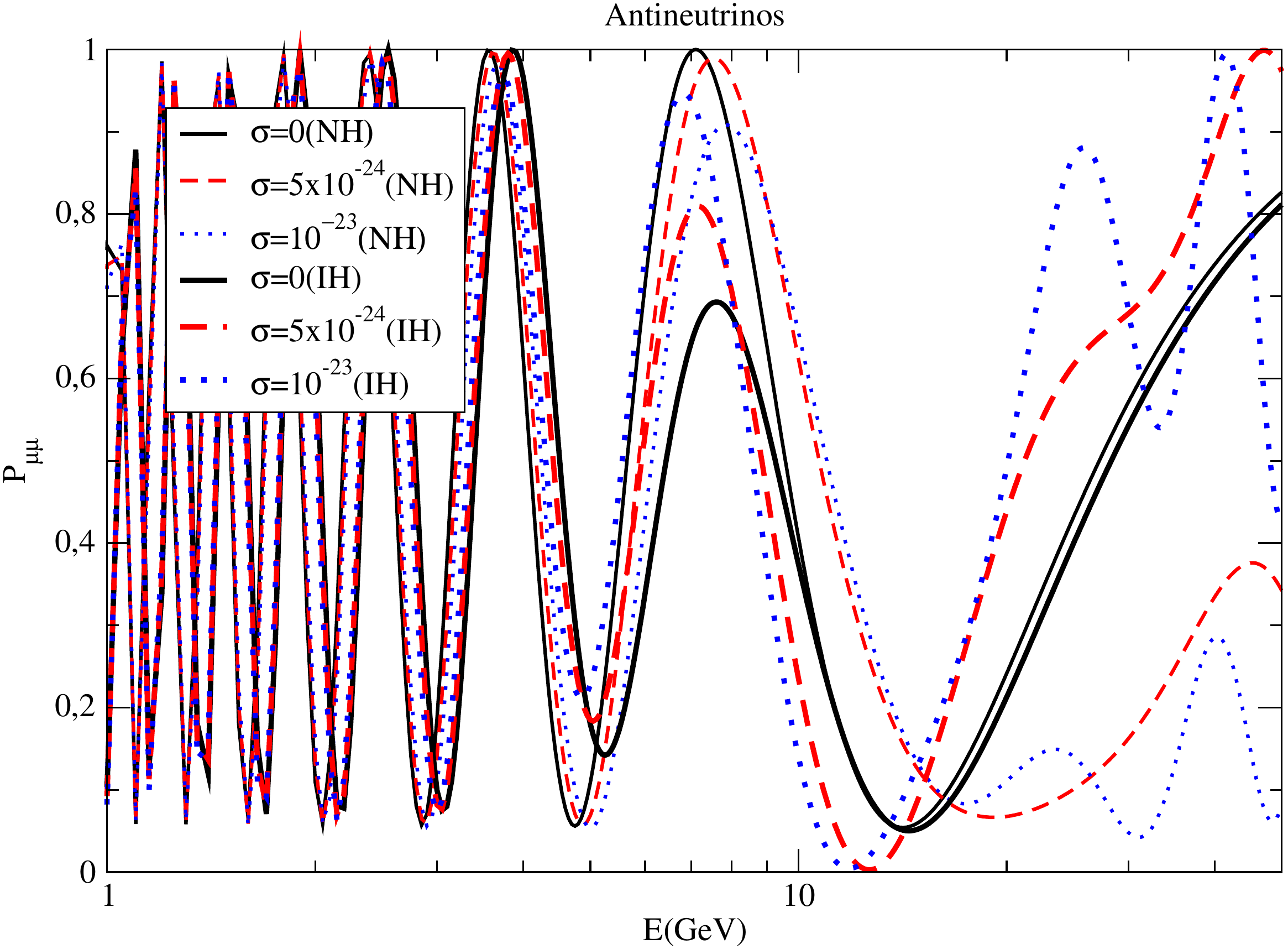}
\caption{In this figure we show the conversion probability for the process $\bar\nu_\mu\to\bar\nu_\mu$. We consider a baseline of $L=7500$~km. The solid curve is for the standard picture ($\sigma=0$) and the dashed and dotted curves are the LIV cases with $\sigma=5\times10^{-24}$ and $\sigma=10^{-23}$, respectively. The thinner curves are for NH and the thicker curves are for IH.}\label{fig4}
\end{figure}

In Fig.~\ref{fig3} and Fig.~\ref{fig4}, respectively, we show the survival probabilities for the processes $\nu_\mu\to\nu_\mu$ and $\bar\nu_\mu\to\bar\nu_\mu$. 
From these figures we infer that the modifications between the LIV case and the standard one occur approximately for $E\ge6$~GeV. Therefore, we deduce that LIV can only be probed for high energy neutrino factories because in this energy regime the main modifications in the oscillation pattern occur.  

\section{Neutrino Detection}\label{detection}

Neutrinos will be detected at a far site of about $L=7500$~km from the neutrino factory. The detection will be done using a magnetized 10~kton iron detector, where we consider and idealize a detection efficiency of 100\%. The detection is done when the $\nu_\mu$ interacts with the nucleus, $N$, and produces the correlated lepton, $\mu^-$. The same happens for $\bar\nu_\mu$, but the lepton associated is the $\mu^+$. The charged-current (CC) cross sections related to these processes are given by~\cite{boehm}
\begin{equation}
\sigma_{\nu N}\approx 0.67\times 10^{-38}\times E(\mbox{GeV})~\mbox{cm}^2
\label{crossnu}
\end{equation}
and
\begin{equation}
\sigma_{\bar\nu N}\approx 0.34\times 10^{-38}\times E(\mbox{GeV})~\mbox{cm}^2.
\label{crossantinu}
\end{equation}
One possible detector being considered is the one called MIND~\cite{mind}. The MIND detector will be able to identify the channel of oscillation by the production of muons and their respective charge signals. An advantage of muon detection is that the backgrounds are very small, of the order of $3\times 10^{-5}$. This is very small if we compare with $\nu_e$ detection in superbeams experiments~\cite{superbeams}. Considering all of this context, we will present our results in the next section with the very optimistic knowledge of backgrounds and a very clean signal.

\section{Results}\label{results}

We consider, also in an optimistic way, a threshold for neutrino detection of 1~GeV. As pointed out before, we take into consideration in the production mechanism $2\times 10^{20}$ muon decays per year that will produce the neutrino spectra characterized by Eq.~(\ref{dist_nu_muon}) and Eq.~(\ref{dist_nu_e}). After the neutrino production, we have a propagation for 7500~km along the Earth, where our $\sigma$ LIV factor is going to be taken into account and we numerically solve Eq.~(\ref{scho}). The detection of these neutrinos is represented by the cross sections in Eqs.~(\ref{crossnu},\ref{crossantinu}) and we use a detector size of 10 kton.   

\begin{figure}[h!tb]
\centering
\includegraphics[scale=0.8]{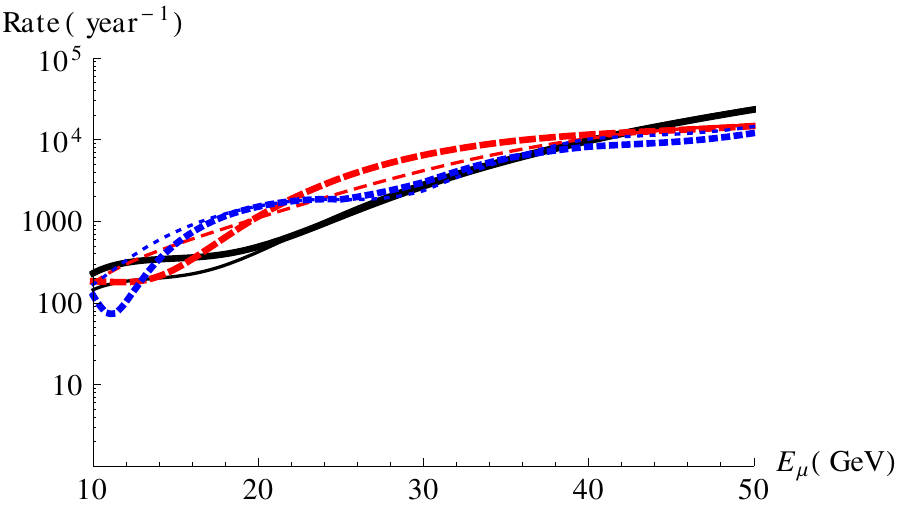}
\caption{The number of events, $\nu_\mu+\bar\nu_\mu$, per year for a 10~kton detector considering neutrinos produced in a neutrino factory with a $\mu^-$ ring for several muon momenta. The legend of our curves respects the same pattern of the other figures and has been omitted.}\label{fig5}
\end{figure}

\begin{figure}[h!tb]
\centering
\includegraphics[scale=0.8]{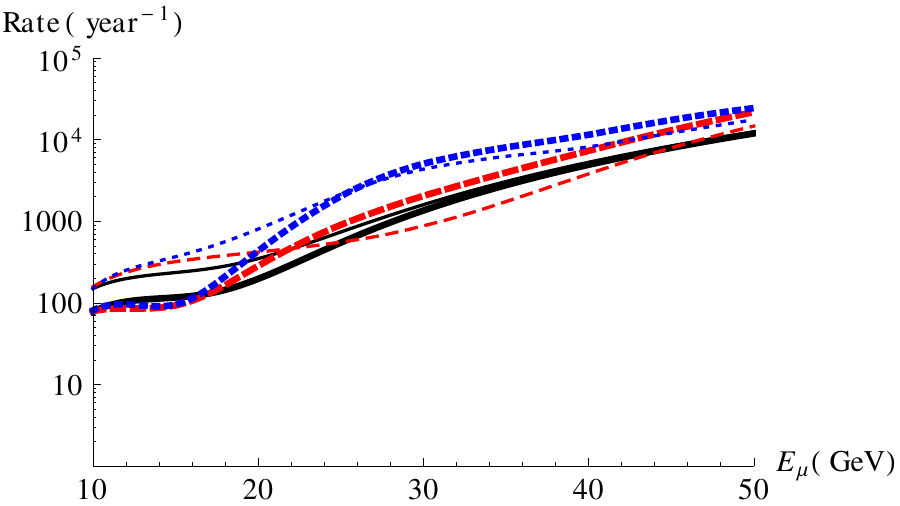}
\caption{The same as Fig.~\ref{fig5}, but for a $\mu^+$ ring.}\label{fig6}
\end{figure}

In Fig.~\ref{fig5}, we show the total number of events ($\nu_\mu+\bar\nu_\mu$) per year considering a $\mu^-$ ring for several muon momenta, which vary from 10 GeV/c to 50 GeV/c. The solid lines are for the standard case, i.e., $\sigma=0$. The red dashed lines are for the LIV case, $\sigma=5\times 10^{-24}$, and the dotted blue lines are for the LIV case, $\sigma=10^{-23}$. The thicker lines represent the IH and the thinner ones the NH. The same is done in Fig.~\ref{fig6}, but considering a $\mu^+$ ring. 

We notice from both figures that we can distinguish the rate if we consider the LIV factor from the standard case. This happens for each muon momentum in the neutrino factory. The main reason for these modifications is the distinct behavior of the oscillations in the LIV cases, as shown in Figs.~\ref{fig1},\ref{fig2},\ref{fig3} and \ref{fig4}, especially for $E>10$~GeV where the modifications become more pronounced. 

To better understand the inclusion of the LIV parameter, we calculated the number of events for two specific momenta of the muon: $p=20$~GeV/c and $p=50$~GeV/c. The calculation has been done for both NH and IH cases and for considering different muon charges ($\mu^\pm$). They are presented in Table~\ref{nevents20} and Table~\ref{nevents50}. In Table~\ref{nevents20}, for both muon polarizations and both hierarchies, we notice an increase in the total number of events ($\nu_\mu+\bar\nu_\mu$) per year when we introduce the LIV factor. First, this increase for the $\mu^-$ ring can be explained by the fact that in both LIV cases, $P_{e\mu}$ and $P_{\mu\mu}$ for $\bar\nu_e\to\bar\nu_\mu$ and $\nu_\mu\to\nu_\mu$, respectively, are greater than $P_{e\mu}$ and $P_{\mu\mu}$ in the non-LIV case. See Fig.~\ref{fig1} and Fig.~\ref{fig4} for regions where $E<20$~GeV. Also, notice that the number of events in the $\mu^-$ ring is larger than in the $\mu^+$ ring. This happens because, for the $\mu^+$ ring, modifications in the probabilities curves are less drastic than in the case of the $\mu^-$ ring for $E<20$~GeV. 

In Table~\ref{nevents50}, we evaluated the number of events ($\nu_\mu+\bar\nu_\mu$) per year for muons with momentum equal to 50 GeV/c. For the $\mu^-$ ring, there is a reduction in the number of events when the $\sigma$ LIV parameter is $\ne 0$, since $P_{\mu\mu}$ for the oscillation channel $\nu_\mu\to\nu_\mu$, is suppressed for neutrino energies higher than 20 GeV. This suppression is even more pronounced when $\sigma=10^{-23}$. However, in the case of a $\mu^+$ ring, generally we can notice an increase in $P_{e\mu}$ for the process $\nu_e\to\nu_\mu$, and $P_{\mu\mu}$ for the channel $\bar\nu_\mu\to\bar\nu_\mu$. This explains the increase in the number of events compared to the non LIV case. The fact that NH has fewer events than IH, for $\sigma=5\times 10^{-24}$ and $\sigma=10^{-23}$, is due to the fact that $P_{\mu\mu}$ (dotted and dashed curves) for $E>15$~GeV, is smaller than $P_{\mu\mu}$ in the $\sigma=0$ case (solid curve), as seen in Fig.~\ref{fig4}.  

\begin{table}[ht]
\centering
\begin{tabular}{ccccc}
\hline
Ring&Hierarchy&$\sigma=0$ & $\sigma=5\times 10^{-24}$ & $\sigma=10^{-23}$ \\
\hline
$\mu^-$&NH & 422 & 1103 & 1533 \\
\hline
$\mu^-$&IH & 488 & 1147 & 1511 \\
\hline
$\mu^+$&NH & 349 & 420 & 803 \\
\hline
$\mu^+$&IH & 197 & 285 & 434 \\
\hline
\end{tabular}
\caption{The number of events, $\nu_\mu+\bar\nu_\mu$, per year for 10~kton detector considering a 7500~km neutrino factory baseline for $\sigma=0$, $\sigma=5\times 10^{-24}$, and $\sigma=10^{-23}$. We consider here NH and IH. Calculations were also done for a $\mu^-$ ring and $\mu^+$ ring with momentum $p=20$~GeV/c.}
\label{nevents20}
\end{table} 

Notice that we have presented our results in terms of the total number of events in one year of data from a neutrino factory. We could have presented our results in terms of the modification of the neutrino spectra, which is related to the neutrino oscillation probabilities, since we notice a severe modification in the pattern of the curves in Fig.~\ref{fig1},~\ref{fig2},~\ref{fig3}, and~\ref{fig4} when we compare to the standard case. However, we think that the strategy of counting the number of events could also be an indication of the presence of LIV phenomena that, combined with the modifications of the spectra, would be a strong indication of new physics.

\begin{table}[ht]
\centering
\begin{tabular}{ccccc}
\hline
Ring&Hierarchy&$\sigma=0$ & $\sigma=5\times 10^{-24}$ & $\sigma=10^{-23}$ \\
\hline
$\mu^-$&NH & 24414 & 15057 & 14492 \\
\hline
$\mu^-$&IH & 23314 & 14681 & 12043 \\
\hline
$\mu^+$&NH & 12414 & 14637 & 17480 \\
\hline
$\mu^+$&IH & 11908 & 21699 & 24310 \\
\hline
\end{tabular}
\caption{The number of events, $\nu_\mu+\bar\nu_\mu$, per year for 10~kton detector considering a 7500~km neutrino factory baseline for $\sigma=0$, $\sigma=5\times 10^{-24}$, and $\sigma=10^{-23}$. We consider here NH and IH. Calculations were also done for a $\mu^-$ ring and $\mu^+$ ring with momentum $p=50$~GeV/c.}
\label{nevents50}
\end{table} 

\section{Conclusions}\label{conclusions}
In this work we analyzed the possible modifications in the number of events ($\nu_\mu+\bar\nu_\mu$) per year in a neutrino factory located 7500~km from a 10~kton detector. We introduced a parameter, $\sigma$, that breaks the Lorentz invariant assumption. This can have consequences in the neutrino oscillation mechanisms and modify the number of events. In Fig.~\ref{fig5} and Fig.~\ref{fig6} we have shown that the modification in the number of events happens when one compares the standard case ($\sigma=0$) to the LIV cases ($\sigma=5\times10^{-24}$ and $\sigma=10^{-23}$) for several muon momenta, both muon polarizations, and NH/IH. These parameter values were chosen considering the baseline and typical neutrino energies of a neutrino factory, since they generate clear modifications in the survival probabilities ($\nu_\mu\to\nu_\mu$ and $\bar\nu_\mu\to\bar\nu_\mu$) and in the conversion probabilities ($\nu_e\to\nu_\mu$ and $\bar\nu_e\to\bar\nu_\mu$). 

If we consider a simplified version of the SME, considering an isotropical approximation (no LIV from rotations), the parameter $\sigma$ of this work is comparable with the parameter $c^{(4)}_{e\mu}$, a dimension 4 operator, that controls neutrino oscillation in the channel of $\nu_e \leftrightarrow \nu_\mu$ and does not promote CPT violation effect. 
Considering data information from LSND and MinibooNE, a maximum sensitivity of 10$^{-19}$ (Table S4 of Ref.~\cite{table_liv}) can be found. For more details of the model and the evaluation of this LIV parameter, see~\cite{newkost}.

In the case of non-terrestrial experiments, in~\cite{atmliv1}, for example, considering the $\nu_\mu \leftrightarrow \nu_\tau$ oscillation channel and the Super-Kamiokande data of atmospheric neutrinos with a range of energy of about four decades, the standard scenario was confirmed as the dominant effect in neutrino oscillation and other terms that induce violations in relativity as subleading effects. With the inclusion of K2K data,~\cite{atmliv2} obtained the same conclusion. Using the Macro data of upward-moving muons of atmospheric neutrinos,~\cite{atmliv3} also found that LIV is not favored by the atmospheric data, even as a subleading term in the oscillation mechanism. At 90\% C.L. these works found, roughly speaking, that $\sigma \lesssim 6\times 10^{-24}$. Despite the fact that atmospheric neutrinos disfavor LIV in the neutrino oscillation context, the results in neutrino factories show that if we have a possible violation in the principle of Lorentz invariance, this can be detected in a future neutrino factory, considering values of $\sigma$ -- $5\times 10^{-24}$ and $10^{-23}$ in this present work -- near of some of the upper limits experimentally determined by the results of atmospheric neutrinos, whose flux is reasonably well understood: normalizations are known to 20\% approximately (10\% approximately for neutrino energies below 10~GeV); the ratio of fluxes are known to $\sim 5\%$; fluxes decay rapidly with neutrino energies of $E_\nu>1$~GeV. So, considering that the known flux properties of atmospheric neutrinos have several uncertainties and the perspective of very clean signals in neutrino factories, neutrino factories in the future can achieve lower uncertainties and reveal the LIV phenomenon or even put more stringent bounds on it considering the neutrino oscillation perspective.

\acknowledgements

The author would like to thank Conselho Nacional de Desenvolvimento Cient\'ifico e Tecnol\'ogico (CNPq) for the financial support. Also, he would like to thank Prof. M. M. Guzzo for reading this material and making very useful suggestions and is grateful to two anonymous referees of J. Phys. G for feedback and excellent recommendations.


\begin{thebibliography}{9}

\bibitem{nufactseminal} 
  S.~Geer,
  Phys.\ Rev.\ D {\bf 57}, 6989 (1998)
  [Erratum-ibid.\ D {\bf 59}, 039903 (1999)]
  [hep-ph/9712290].

\bibitem{nufact1} 
Y.~Kuno, Y.~Mori, S.~Machida, T.~Yokoi, Y.~Iwashita, J.~Sato and O.~Yasuda,
  NUFACTJ-05-24-2001.

\bibitem{nufact2}
B.~Autin, A.~Blondel and J.~R.~Ellis,
  CERN-99-02, CERN-YELLOW-99-02.
  
\bibitem{nufact3}
D.~Finley and N.~Holtkamp,
  Nucl.\ Instrum.\ Meth.\ A {\bf 472}, 388 (2000).

\bibitem{nufact4} 
S.~Ozaki, R.~B.~Palmer, M.~S.~Zisman, J.~C.~Gallardo, M.~Goodman, A.~Hassanein, J.~H.~Norem and C.~B.~Reed {\it et al.},
  BNL-52623, FERMILAB-PUB-01-544-A.

\bibitem{nufact5}
P.~Gruber, M.~Aleksa, J.~F.~Amand, B.~Autin, J.~L.~Baldy, M.~Benedikt, R.~Bennett and A.~Bernadon {\it et al.},
  CERN Yellow Report CERN-2004-002, pp.7-86

\bibitem{nufact6}
M.~M.~Alsharoa {\it et al.}  [Muon Collider/Neutrino Factory Collaboration],
  Phys.\ Rev.\ ST Accel.\ Beams {\bf 6}, 081001 (2003)
  [hep-ex/0207031].

\bibitem{nufactinterest1} 
  A.~Stahl, C.~Wiebusch, A.~M.~Guler, M.~Kamiscioglu, R.~Sever, A.~U.~Yilmazer, C.~Gunes and D.~Yilmaz {\it et al.},
  CERN-SPSC-2012-021.

\bibitem{nufactinterest2} 
  A.~Bandyopadhyay {\it et al.}  [ISS Physics Working Group Collaboration],
  Rept.\ Prog.\ Phys.\  {\bf 72}, 106201 (2009)
  [arXiv:0710.4947 [hep-ph]].

\bibitem{perspective_lorentz1} 
  B.~-Q.~Ma,
  arXiv:1203.5852 [hep-ph].

\bibitem{perspective_lorentz2} 
  B.~-Q.~Ma,
  Int.\ J.\ Mod.\ Phys.\ Conf.\ Ser.\  {\bf 10}, 195 (2012)
  [arXiv:1203.0086 [hep-ph]].

\bibitem{perspective_lorentz3} 
  S.~Liberati,
  Class.\ Quant.\ Grav.\  {\bf 30}, 133001 (2013)
  [arXiv:1304.5795 [gr-qc]].

\bibitem{table_liv} 
  V.~A.~Kostelecky and N.~Russell,
  Rev.\ Mod.\ Phys.\  {\bf 83}, 11 (2011)
  [arXiv:0801.0287 [hep-ph]].

\bibitem{sme1} 
  D.~Colladay and V.~A.~Kostelecky,
  Phys.\ Rev.\ D {\bf 55}, 6760 (1997)
  [hep-ph/9703464].

\bibitem{sme2} 
  D.~Colladay and V.~A.~Kostelecky,
  Phys.\ Rev.\ D {\bf 58}, 116002 (1998)
  [hep-ph/9809521].

\bibitem{sme3} 
  J.~S.~Diaz and V.~A.~Kostelecky,
  Phys.\ Lett.\ B {\bf 700}, 25 (2011)
  [arXiv:1012.5985 [hep-ph]].

\bibitem{sme4} 
  J.~S.~Diaz and A.~Kostelecky,
  Phys.\ Rev.\ D {\bf 85}, 016013 (2012)
  [arXiv:1108.1799 [hep-ph]].

\bibitem{neutrinos_lorentz1} 
  V.~A.~Kostelecky and M.~Mewes,
  Phys.\ Rev.\ D {\bf 69}, 016005 (2004)
  [hep-ph/0309025].

\bibitem{neutrinos_lorentz2} 
  J.~S.~Diaz,
  arXiv:1109.4620 [hep-ph].

\bibitem{neutrinos_lorentz3} 
  A.~Kostelecky and M.~Mewes,
  Phys.\ Rev.\ D {\bf 85}, 096005 (2012)
  [arXiv:1112.6395 [hep-ph]].

\bibitem{neutrinos_lorentz4} 
  J.~S.~Diaz,
  Adv.\ High Energy Phys.\  {\bf 2014}, 962410 (2014)
  [arXiv:1406.6838 [hep-ph]].

\bibitem{minos} 
  P.~Adamson {\it et al.}  [MINOS Collaboration],
  Phys.\ Rev.\ D {\bf 76}, 072005 (2007)
  [arXiv:0706.0437 [hep-ex]].

\bibitem{opera} 
  T.~Adam {\it et al.}  [OPERA Collaboration],
  JHEP {\bf 1301}, 153 (2013)
  [arXiv:1212.1276 [hep-ex]].

\bibitem{icarus} 
  M.~Antonello, B.~Baibussinov, P.~Benetti, F.~Boffelli, E.~Calligarich, N.~Canci, S.~Centro and A.~Cesana {\it et al.},
  JHEP {\bf 1211}, 049 (2012)
  [arXiv:1208.2629 [hep-ex]].

\bibitem{lvd} 
  N.~Y.~.Agafonova {\it et al.}  [LVD Collaboration],
  Phys.\ Rev.\ Lett.\  {\bf 109}, 070801 (2012)
  [arXiv:1208.1392 [hep-ex]].

\bibitem{coleman} 
  S.~R.~Coleman and S.~L.~Glashow,
  Phys.\ Rev.\ D {\bf 59}, 116008 (1999)
  [hep-ph/9812418].

\bibitem{cpt1} 
  S.~M.~Bilenky, M.~Freund, M.~Lindner, T.~Ohlsson and W.~Winter,
  Phys.\ Rev.\ D {\bf 65}, 073024 (2002)
  [hep-ph/0112226].

\bibitem{cpt2} 
  J.~S.~Diaz, V.~A.~Kostelecky and M.~Mewes,
  Phys.\ Rev.\ D {\bf 80}, 076007 (2009)
  [arXiv:0908.1401 [hep-ph]].


\bibitem{geer2} 
  S.~Geer,
  arXiv:1202.2140 [physics.acc-ph].

\bibitem{gaisser} 
  T.~K.~Gaisser,
  ``Cosmic rays and particle physics,''
  Cambridge, UK: Univ. Pr. (1990).

\bibitem{barger} 
  V.~Barger, D.~Marfatia and K.~Whisnant,
  ``The physics of neutrinos,''
  Princeton University Press (2012).

\bibitem{tunnell} 
  C.~D.~Tunnell, J.~H.~Cobb and A.~D.~Bross,
  arXiv:1111.6550 [hep-ph].

\bibitem{winter} 
  W.~Winter,
  Phys.\ Rev.\ D {\bf 85}, 113005 (2012)
  [arXiv:1204.2671 [hep-ph]].

\bibitem{china_proposition} 
  J.~Cao, M.~He, Z.~L.~Hou, H.~T.~Jing, Y.~F.~Li, Z.~H.~Li, Y.~P.~Song and J.~Y.~Tang {\it et al.},
  Phys.\ Rev.\ ST Accel.\ Beams {\bf 17}, no. 9, 090101 (2014)
  [Phys.\ Rev.\ S.\ T.\ A.\ B.\  {\bf 17}, 090101 (2014)]
  [arXiv:1401.8125 [physics.acc-ph]].

\bibitem{msw}
  L.~Wolfenstein,
  Phys.\ Rev.\ D {\bf 17}, 2369 (1978);
S.~P.~Mikheev and A.~Y.~.Smirnov,
  Sov.\ J.\ Nucl.\ Phys.\  {\bf 42}, 913 (1985)
  [Yad.\ Fiz.\  {\bf 42}, 1441 (1985)].

\bibitem{magic} 
  P.~Huber and W.~Winter,
  Phys.\ Rev.\ D {\bf 68}, 037301 (2003)
  [hep-ph/0301257].

\bibitem{freund1} 
  M.~Freund, M.~Lindner, S.~T.~Petcov and A.~Romanino,
  Nucl.\ Phys.\ B {\bf 578}, 27 (2000)
  [hep-ph/9912457].

\bibitem{freund2} 
  M.~Freund, P.~Huber and M.~Lindner,
  Nucl.\ Phys.\ B {\bf 585}, 105 (2000)
  [hep-ph/0004085].


\bibitem{prem} 
  A.~M.~Dziewonski and D.~L.~Anderson,
  Phys.\ Earth Planet.\ Interiors {\bf 25}, 297 (1981).

\bibitem{con1} 
  P.~I.~Krastev and S.~T.~Petcov,
  Phys.\ Lett.\ B {\bf 205}, 84 (1988).

\bibitem{con2} 
  M.~Chizhov, M.~Maris and S.~T.~Petcov,
  hep-ph/9810501.

\bibitem{con3} 
  S.~T.~Petcov,
  Phys.\ Lett.\ B {\bf 434}, 321 (1998)
  [hep-ph/9805262].

\bibitem{earth_profile} 
  I.~Mocioiu and R.~Shrock,
  Phys.\ Rev.\ D {\bf 62}, 053017 (2000)
  [hep-ph/0002149].

\bibitem{greenberg} 
  O.~W.~Greenberg,
  Phys.\ Rev.\ Lett.\  {\bf 89}, 231602 (2002)
  [hep-ph/0201258].

\bibitem{newkost} 
  A.~Kostelecky and M.~Mewes,
  Phys.\ Rev.\ D {\bf 85}, 096005 (2012)
  [arXiv:1112.6395 [hep-ph]].

\bibitem{cp_violation_ignore} 
  V.~Barger, D.~Marfatia and K.~Whisnant,
  Phys.\ Rev.\ D {\bf 65}, 073023 (2002)
  [hep-ph/0112119].

\bibitem{bari} 
  G.~L.~Fogli, E.~Lisi, A.~Marrone, D.~Montanino, A.~Palazzo and A.~M.~Rotunno,
  Phys.\ Rev.\ D {\bf 86}, 013012 (2012)
  [arXiv:1205.5254 [hep-ph]].

\bibitem{boehm} 
  F.~Boehm and P.~Vogel,
  ``Physics Of Massive Neutrinos,''
  CAMBRIDGE, UK: UNIV. PR. (1987).

\bibitem{mind} 
  R.~Bayes, A.~Bross, A.~Cervera, M.~Ellis, A.~Laing, F.~J.~P.~Soler and R.~Wands,
  J.\ Phys.\ Conf.\ Ser.\  {\bf 408}, 012075 (2013).

\bibitem{superbeams} 
  V.~D.~Barger, S.~Geer, R.~Raja and K.~Whisnant,
  Phys.\ Rev.\ D {\bf 63}, 113011 (2001)
  [hep-ph/0012017].



\bibitem{atmliv1} 
  G.~L.~Fogli, E.~Lisi, A.~Marrone and G.~Scioscia,
  Phys.\ Rev.\ D {\bf 60}, 053006 (1999)
  [hep-ph/9904248].

\bibitem{atmliv2} 
  M.~C.~Gonzalez-Garcia and M.~Maltoni,
  Phys.\ Rev.\ D {\bf 70}, 033010 (2004)
  [hep-ph/0404085].

\bibitem{atmliv3} 
  G.~Battistoni, Y.~Becherini, S.~Cecchini, M.~Cozzi, H.~Dekhissi, L.~S.~Esposito, G.~Giacomelli and M.~Giorgini {\it et al.},
  Phys.\ Lett.\ B {\bf 615}, 14 (2005)
  [hep-ex/0503015].

\end{thebibliography}
\end{document}